\documentclass[12pt,english]{article}
\usepackage[T1]{fontenc}
\usepackage[latin9]{inputenc}
\usepackage{setspace}
\makeatletter
\usepackage{aStyle}

\makeatother

\usepackage{babel}
\begin{document}
\begin{doublespace}
\begin{center}
\textbf{\Large{}The Epistemology of Contemporary Physics:\\Classical
Mechanics I}\vspace{-1.3cm}
\par\end{center}
\end{doublespace}

\begin{center}
Taha Sochi (Contact: ResearchGate)\vspace{-0.4cm}
\par\end{center}

\begin{center}
London, United Kingdom
\par\end{center}

\noindent \textbf{Abstract}: In this paper of ``The Epistemology
of Contemporary Physics'' series we investigate the epistemological
significance and sensibility (and hence interpretability and interpretation)
of classical mechanics in its Newtonian and non-Newtonian formulations.
As we will see, none of these formulations provide a clear and consistent
framework for understanding the physics which they represent and hence
they all represent valid formalism without proper epistemology or
sensible interpretation.\vspace{0.3cm}


\clearpage{}

\tableofcontents{}

\clearpage{}

\section{Introduction}

Classical mechanics is one of the most successful and useful scientific
theories ever created despite its limitations due largely to its restricted
domain of validity (i.e. classical macroscopic scale and inertial
frames of reference).\footnote{In fact, there are other limitations and restrictions on the domain
of validity (whether practically or conceptually) of classical mechanics
or/and some of its formulations. For instance, the Newtonian formulation
of classical mechanics is essentially about the mechanics of particles.} It is tested successfully on countless terrestrial and celestial
physical systems of different magnitudes and various specific features
in observation, experimentation and application. In fact, it is the
main mechanical theory used in scientific investigations and engineering
applications at the classical macroscopic scale (noting the fact that
it may be considered as a good approximation in some of its topics
and aspects).

There are several formulations to classical mechanics such as Newtonian,
Lagrangian and Hamiltonian (noting that the existence of more than
one formulation of classical mechanics represents a demonstration
of the principle of non-uniqueness of science and a simple instance
of it; see $\S$ 5.3 of \cite{SochiEpistIntro2024}). The dominant
of these formulations is the Newtonian formulation which is represented
mainly by Newton's three laws of motion, i.e. the law of inertia,
the law of momentum or acceleration and the law of interaction (see
for instance \cite{TaylorBook2005,HallidayRWBook2014}).

In this paper we briefly investigate the main epistemological characteristics
of the Newtonian formulation of classical mechanics (and its interpretability
and interpretation in particular). A similar (but much shorter) investigation
will be conducted on the non-Newtonian formulations of classical mechanics
(considering essentially the main two versions of these formulations;
namely the Lagrangian and Hamiltonian mechanics).

However, before we start our investigation it is worth noting the
following points:
\begin{enumerate}
\item ``Classical mechanics'' (which is in the title of this paper) has
different meanings in the literature depending on the authors and
contexts, e.g. it can be contrasted to quantum mechanics or to the
relativity theories (or to other physical theories and branches).
What we mean here by ``classical mechanics'' is the mechanics prior
to the emergence of the new branches of mechanics of modern physics
such as quantum mechanics and Lorentz mechanics.
\item The focus of the present paper is actually the ``laws of motion''
part of classical mechanics (noting that classical mechanics, as defined
already, includes other topics and branches such as gravity).\footnote{``The laws of motion'' should be understood as more general than
Newton's laws of motion because we are considering non-Newtonian formulations
(as well as Newtonian formulation). So, it should be understood to
mean something like ``the mechanics of motion''. For more clarity,
the investigation of this paper is about the Newtonian mechanics (represented
by Newton's three laws of motion) and the Lagrangian and Hamiltonian
mechanics (representing the main formulations of non-Newtonian classical
mechanics).}
\item The focal point of the investigation of the present paper is the epistemological
sensibility and interpretability (and interpretation) of classical
mechanics (noting that other epistemological aspects and topics of
classical mechanics should be investigated in upcoming papers).
\end{enumerate}
\vspace{0.2cm}The structure of this paper is that we investigate
in the next two sections the Newtonian formulation and non-Newtonian
formulations of classical mechanics. We then conclude our investigation
with a ``Conclusions'' section in which we summarize the main facts
and results that we discussed and obtained in this paper. Our plan
is to investigate other epistemological aspects of classical mechanics
(which are more specific or not related directly to interpretation
and interpretability) in forthcoming papers of this series (as indicated
already).

\section{Newtonian Formulation of Classical Mechanics}

The Newtonian formulation of classical mechanics is the most common
and the most intuitive formulation of classical mechanics. Moreover,
it historically precedes all other formulations of classical mechanics
and hence it is the mother of all formulations of classical mechanics
(noting that classical mechanics is the mother of all modern physics
because virtually all the branches and theories of modern physics
started from or based on classical mechanical deliberations and considerations).

In this section we investigate (following some preliminary reflections
on Newton's laws of motion) the main features of this formulation
and its main limitations and shortcomings from an epistemological
perspective (or largely so) to reach a conclusion that this formulation
has no sensible, consistent and complete epistemology and hence it
has no valid interpretation although in principle it has the capability
to be interpretable due to its rich, diverse and intuitive conceptual,
philosophical and epistemological framework.

\subsection{Preliminary Reflections on Newton's Laws of Motion}

Before we go through our main investigation in this section (related
to the features and limitations of the Newtonian formulation) it is
useful to be aware of the following preliminary reflections on the
Newtonian formulation as represented mainly by Newton's laws of motion.
These reflections are outlined in the following subsections.

\subsubsection{\label{subsecRelationNewton12}Relationship between Newton's First
and Second Laws}

It is common to consider Newton's first law as a special case of Newton's
second law. Although this is not incorrect it is partially reflecting
the actual relationship between Newton's first and second laws and
hence it can be misleading. The reality is that Newton's first law
is partially included in Newton's second law in the sense that Newton's
first law contains more content and substance than what is implied
by Newton's second law. To be more clear, Newton's first law has two
ingredients:
\begin{enumerate}
\item The proposition that: absence of force results in absence of acceleration
(or absence of temporal change of linear momentum). This ingredient
is implied by Newton's second law\footnote{Actually, from a purely formal and logical perspective what is implied
by Newton's second law is less than this because the causality of
the absence of force to the absence of acceleration (which is implied
by the statement ``absence of force results in absence of acceleration'')
is not incorporated within Newton's second law (as will be investigated
partly later on). So, what is actually incorporated is the association
of force and acceleration in the sense that the presence/absence of
one implies the presence/absence of the other.} and hence from the perspective of this ingredient Newton's first
law is a special case of Newton's second law since this ingredient
of the first law is implied by the second law.
\item The proposition that: absence of force results in having an ``inertial
state of rest or uniform motion''. This ingredient is not implied
by Newton's second law because this particular ``inertial state of
rest or uniform motion'' is more specific than the state of ``absence
of acceleration'' or ``absence of temporal change of linear momentum''.
For example, we can imagine that the absence of force results in a
state of sudden and instant rest where the unaccelerated object (due
to the cessation of force) comes to a state of standstill immediately
and abruptly as soon as force ceases to exist. It should be obvious
that this situation or scenario satisfies Newton's second law (since
absence of force leads to absence of acceleration or absence of change
of momentum) but does not satisfy Newton's first law (since absence
of force does not lead to the aforementioned ``inertial state of
rest or uniform motion'').
\end{enumerate}
\vspace{0.15cm}By the benefit of a deeper insight we can claim that
the essence of all this is the paradigm of inertia which is the actual
essence and substance of Newton's first law and this essence is what
makes Newton's first law more informative than Newton's second law
from this perspective (and hence makes Newton's first law distinct
from Newton's second law despite being partially included in Newton's
second law).

In fact, the paradigm of inertia is pivotal and central to the Newtonian
formulation of classical mechanics (and possibly to classical mechanics
in general) and it has very important implications and consequences
(both formally and epistemologically). For instance, the aforementioned
scenario (of abrupt cessation of motion at the instant of cessation
of force) implies violation of the principles of conservation of energy
and momentum and hence we can claim that the conservation of energy
and momentum are the result of the paradigm of inertia (although being
so should be partially since the conservation of energy and momentum
require more than the paradigm of inertia).

This should make Newton's first law (with its pivotal essence of ``inertia'')
as important as\footnote{And actually ``as independent as'' despite being partially included
in Newton's second law.} Newton's second and third laws (noting that Newton's first law is
usually trivialized in the literature by depicting it as being just
a special case of Newton's second law and totally embedded in it and
hence it is no more than an important special case or a clarifying
statement or attachment to Newton's second law and the Newtonian formulation
of classical mechanics).

However, this trivialization of Newton's first law may be justified
(partially) by noting that the literature of Newton's laws of motion
(and classical mechanics in general) is usually and largely about
the formalism of these subjects; and the paradigm of inertia does
not demonstrate its real power and strength in the formalism (because
it is implicitly and intrinsically embedded within the formalism).
This is unlike the epistemology of these subjects where the paradigm
of inertia is pivotal and central and hence we may say (laxly) that
although Newtons first law is no more than a special case of Newton's
second law from the perspective of formalism, it is more than a special
case of Newton's second law from the perspective of epistemology (due
to the special importance of ``inertia'' in the epistemology but
not in the formalism).

Anyway, any potential disputes about these issues should not be of
substantial value or importance to our current investigation and discussion
and hence we do not pursue this debate further in this paper (although
we may come back to some of these issues in forthcoming papers).

\subsubsection{``Hidden'' Ingredients of Newton's Second Law}

Newton's second law as commonly stated and formulated (i.e. force
equals\footnote{We use ``equals'' for simplicity and brevity. To be more precise
we should use ``is proportional to''.} mass times acceleration or force equals time derivative of linear
momentum, that is $\mathbf{F}=m\mathbf{a}$ or $\mathbf{F}=d\mathbf{p}/dt$)
misses an important ingredient of this law (especially from an epistemological
perspective) which is the ``implicit'' fact that force is the cause
of acceleration or change of linear momentum. In fact, this causality
relationship between force and acceleration (or change of momentum)
is usually not given sufficient attention (if given any attention
at all) in the literature of physics and textbooks. Again, the reason
seems to be its philosophical and epistemological nature which the
literature of physics and textbooks have no much interest in since
they are generally about the formalism of physics rather than its
philosophy and epistemology.

Now, according to the common understanding among physicists (which
may also be seen as being intuitive and commonsense), force is the
cause of acceleration (as stated already). However, there seems to
be an opposite view that considers acceleration as the cause of force
where this view is mainly based on and justified by the Machian proposal
(i.e. inertia is caused by the overall matter distribution in the
Universe) in the sense that acceleration relative to the rest of the
Universe induces inertial forces in the accelerated massive object.
But whether (according to this view or according to its rationale)
the induced force is the action or reaction force requires further
analysis and clarification.

Anyway, any claim that independently-generated acceleration (i.e.
without motivating force) causes force seems to be nonsensical and
completely in conflict with our intuition which is based on and derived
from our daily life experiences where we (by our free will and conscious
determination) generate forces to produce accelerations (not the other
way around), and hence for the aforementioned view to be logical (or
at least not contradicting our intuition) the force that is supposedly
caused by acceleration should be the reaction (and dependent) force,
i.e. not the action (and independent) force. These issues (as well
as other similar issues) should be the subject of further investigation
and analysis in the future.

It is also useful and appropriate to mention here that there are other
``hidden'' or ``implicit'' facts which are usually not indicated
or noticed in the common statements of Newton's second law and hence
they should be highlighted and remembered. For example, ``force''
in the statement of Newton's second law is the resultant (or net)
force. The obvious implication of this is that force is a necessary
but not sufficient condition for generating acceleration (or change
of momentum), and hence what is necessary and sufficient (within this
context) for generating acceleration is ``unopposed force'' or ``unbalanced
force''.

We should also note that the statement of Newton's second law as ``force
equals mass times acceleration'' (i.e. $\mathbf{F}=m\mathbf{a}$)
is restricted to the cases where mass is constant, and hence the more
rigorous and general statement of Newton's second law is ``force
equals time derivative of linear momentum'' (i.e. $\mathbf{F}=d\mathbf{p}/dt$).

Simple and (rather) delicate details like these should always be noticed
and remembered when analyzing Newton's second law and extracting its
implications and consequences to avoid confusion and blunders. In
fact, this should apply to all other laws and facts.

\subsubsection{\label{subsecNatureNewton2}Nature of Newton's Second Law}

Whether Newton's second law is essentially a definition or a physical
law (or physical fact) is an issue that is usually discussed in the
literature.\footnote{In fact, this discussion may extend occasionally to Newton's laws
of motion in general (e.g. whether the first law is a definition for
inertia or inertial frame).} Our view is that Newton's second law (like any other law) is a physical
law that is based on definition (or rather definitions), and this
is actually the essence of all human knowledge, i.e. human knowledge
is actually an elaborate linguistic system that is built and based
on a collection of definitions and conventions and hence it is creative
and inventive in part and reflective and suggestive in another part.\footnote{These issues are discussed in detail in chapter 2 of \cite{SochiKetab1}.}
So, in this regard Newton's second law is not different from any other
physical law or factual statement or any other piece of human knowledge.

Yes, there is a specific reason that makes Newton's second law more
eligible for this type of investigation and questioning, which is
what we will discuss later on (see $\S$ \ref{subsecLackOfDefinition})
about the lack of technical definition to basic concepts in the Newtonian
formulation (including the concept of force). So, from this perspective
the consideration of Newton's second law as a definition (or rather
technical definition) seems to be motivated by the desire to regard
this law as a convention or definition so that the problem of lack
of technical definition is supposedly addressed. However, as we will
see (refer to $\S$ \ref{subsecCentralityForce}) this way of technically
defining basic and elementary concepts (such as defining force by
Newton's second law) is rather circular and hence this problem is
not addressed properly by this attempt.

We should also note that considering Newton's second law as a definition
may also be justified by the claim that without a definite physical
force law\footnote{In fact, we may need more than one definite physical force law to
identify all the forces in the given physical situation (as discussed
extensively in the textbooks of classical mechanics).} (that identifies the force in Newton's second law) the second law
is physically useless and without any real physical content or substance.
However, although the need for a definite physical force law may partially
legitimize and justify this consideration, the aforementioned claim
is not free of exaggeration since Newton's second law as it is (i.e.
without a definite physical force law) still has real physical content
and significance (especially at the epistemological level) although
it may not be very useful formally and practically in investigating
and analyzing specific physical situations and circumstances (which
are represented typically in solving specific problems in classical
mechanics).

Anyway, if we ignore all these technicalities (and possibly other
similar technicalities) then we can consider Newton's second law as
being both a definition and a physical law at the same time (especially
if we adopt the aforementioned view about human knowledge). In fact,
any specific consideration should depend, in part, on how we build
and construct our conceptual and axiomatic framework for the Newtonian
formulation (e.g. with which collection of primary concepts we start
and how we define or hypothesize them and so on) and the second law
in particular. It should also depend on our adopted philosophical
and epistemological views and opinions (or our doctrine in these regards).
In brief, there is no single correct view (i.e. it is not a black-and-white
issue) since there are many considerations that can determine and
justify (or falsify) the chosen stand about this issue. In fact, there
are many aspects about this issue that deserve to be investigated
and inspected closely; however our investigation so far is enough
for what we need in the immediate future (noting that we will pursue
this investigation further when we need).

\subsubsection{Importance of Newton's Third Law}

The association of Newton's third law with the principles of conservation
of linear and angular momentum is well known, and hence this law enjoys
(from this perspective) an exceptional importance among Newton's laws
of motion noting that the conservation of momentum (which Newton's
third law represents within classical mechanics) is one of the most
fundamental principles in all physics not only in classical mechanics
(and noting as well that Newton's first and second laws do not enjoy
such a distinction by having a similar association with or representation
of a fundamental principle of physics like the conservation of momentum).\footnote{We should also note that Newton's first and second laws may be regarded
as definitions.}

In fact, it may be claimed (from another perspective) that Newton's
third law is the most important law among Newton's three laws of motion
\textit{from a real physical perspective and content} because the
first law is included in the second law while the second law is essentially
a definition without physical content (i.e. it is like an empty shell
without substance inside) because without a definite physical force
law (that identifies the force in Newton's second law) the second
law is physically useless and without any real physical content or
substance; and this (according to this claim) is not the case with
Newton's third law because of its implication of the conservation
of momentum (which is one of the most fundamental and verified principles
of Nature).

However, this claim (apart from being questionable from some of its
bases and foundations such as the claim that the first law is included
in the second law and the claim that the second law is essentially
a definition without physical content which were discussed and questioned
in the previous subsections; see $\S$ \ref{subsecRelationNewton12}
and $\S$ \ref{subsecNatureNewton2}) is not free of exaggeration
because every one of these laws has its role and importance within
the Newtonian formulation and classical mechanics. We should also
mention in this context the potential violations of Newton's third
law (see $\S$ \ref{subsecViolationThirdLaw}) which should cast a
shadow on this claim (and possibly even on the aforementioned exceptional
importance of Newton's third law). Anyway, these issues should be
investigated in detail in the forthcoming papers of this series.

\subsection{Features of Newtonian Formulation of Classical Mechanics}

The main epistemological features (which are largely related to interpretability
and interpretation) of the Newtonian formulation of classical mechanics
are investigated briefly in the following subsections (noting that
some of these features are related to the formalism of the Newtonian
formulation as well although this is of no interest to us in this
investigation whose focus is epistemology). More specific and detailed
investigations to some epistemological features and aspects of this
formulation should be pursued in upcoming papers of this series.

\subsubsection{\label{subsecConceptualFramework}Rich and Intuitive Conceptual Framework}

The Newtonian formulation of classical mechanics is associated with
and based on a rich (and almost complete) conceptual, philosophical
and epistemological framework. A close inspection to the Newtonian
formulation should reveal that this formulation includes (explicitly
or implicitly) almost all the conceptual elements required for the
description of motion (which is the essence of mechanics as defined
to be the science of motion and its causes) kinematically and dynamically
in a clear and deterministic way (e.g. absolute space, absolute time,
matter in its characteristic and quantitative qualification as mass,
causes or agents of motion which are commonly labeled as ``forces'',
etc.).

Furthermore, almost all the basic concepts and elements in the Newtonian
formulation (i.e. mass, force, velocity, acceleration, space, time,
etc.) are intuitive (or virtually intuitive),\footnote{We should admit that our intuition about these concepts (or at least
some of them) should be partly attributed to the formal education
(at elementary and higher levels). However, this does not change the
fact that we have such intuition about these concepts. We should also
note that even the non-intuitive (or less intuitive) concepts in the
Newtonian formulation are generally defined by using intuitive (or
more intuitive) concepts (such as defining the non-intuitive concept
of momentum as the product of mass times velocity which are intuitive
concepts in general).} and this should justify the general feeling (or impression) that
the Newtonian formulation is the mechanics of commonsense, especially
when noting this in its proper historical context and perspective
where this mechanics was conceived and born in the traditional philosophical
environment of Renaissance (or just post Renaissance) which was still
under the strong influence of the ancient or classic ``Aristotelian''
philosophy which is the ``philosophy of commonsense'' (at least
within its ``Natural Philosophy'' part).

Hence, this formulation possesses a proper conceptual framework to
be \textit{interpretable in principle} and even \textit{actually-interpreted}
(although this does not mean that it actually has a proper interpretation).
This is in contrast to (at least some of the) other formulations of
classical mechanics (i.e. the non-Newtonian formulations which will
be investigated in $\S$ \ref{secNonNewtonian}) which lack a proper
conceptual framework and hence they are not interpretable as we will
see later on.

It is worth noting the following points about the Newtonian formulation
and its conceptual, philosophical and epistemological framework (noting
that some of these points apply to classical mechanics in general):
\begin{enumerate}
\item \label{enuIntrinsicExtrinsic}We can identify two main parts of the
conceptual framework of the Newtonian formulation: an intrinsic part
which is presumed within (or based on or implied by or ... etc.) the
formalism of this formulation and an extrinsic (or appended) part
which is attached historically to this formulation and it is consistent
and compliant with the formalism and its spirit although it is not
necessarily required by (or based on or implied by or ... etc.) the
formalism. In general, we do not distinguish between these two parts
although the intrinsic part is the most significant epistemologically
(and hence the most important to us).
\item We can say that the Newtonian formulation (within its conceptual framework)
is characterized by being \textit{absolute} and \textit{deterministic}
in the sense that all the physical quantities and relationships in
this formulation are defined and quantified in an absolute sense and
hence all the physical quantities can \textit{in principle} be determined
quantitatively with an infinite precision while all the outcomes of
the events and occurrences can be completely determined \textit{in
principle} according to the given conditions and circumstances (and
regardless of any observer or frame of reference). So, any ambiguity
or uncertainty about the quantities and outcomes should be caused
by casual ignorance of the observer and not by intrinsic indetermination
of the factors involved in the presumed physical situation. In fact,
this represents a strong form of realism where the observer and the
observed are totally independent of each other and where the observed
has complete and definite reality (irrespective of the observer or
frame of reference).\\
It should be obvious that being \textit{absolute} and \textit{deterministic}
should provide the Newtonian formulation with complete clarity, sensibility
and intuitivity both in formalism as well as in epistemology, and
hence these characteristics should be seen as an advantage in this
formulation (and in its epistemology in particular) although this
does not mean that these characteristics are free of problems, question
marks and criticisms.
\item We can say that the Newtonian formulation (within its conceptual framework)
is consistent in general with all the (relevant)\footnote{We note that even the principle of non-uniqueness of science is realized
within the classical mechanics which the Newtonian formulation represents
one of its variants (as indicated earlier).} epistemological principles of science (see $\S$ 5 of \cite{SochiEpistIntro2024}).
The principles of reality and truth as well as the principle of causality
are obviously satisfied within the conceptual framework of this formulation
(at least by not having contradictions between these principles and
this formulation). This should also be the case with regard to the
principle of intuitivity (as declared earlier that this conceptual
framework is generally intuitive). The principle of economy (noting
that it is not a necessity or obligation) may also be satisfied by
comparing this formulation to other formulations of classical mechanics
(and even to other types of mechanics); moreover intuitivity should
imply economy in some sense (see $\S$ 2.4.5 of \cite{SochiBook8}).
Anyway, this formulation is at least not in conflict with the principle
of economy (noting that the Newtonian formulation is not unnecessarily
excessive or elaborate in its formalism and epistemology) and this
should be enough for meeting the criterion of being \textit{consistent}
with the principle of economy (as well as the other relevant epistemological
principles of science).\\
It should be obvious that being consistent with the epistemological
principles of science should provide the Newtonian formulation with
complete ``epistemological legitimacy'' from this perspective, and
hence this characteristic should be seen as another advantage in this
formulation.
\end{enumerate}

\subsubsection{\label{subsecAbsoluteFrame}Existence of Absolute Frame of Reference}

This is an obvious epistemological feature that characterizes the
Newtonian formulation of classical mechanics (noting that the main
non-Newtonian formulations of classical mechanics lack a proper conceptual,
philosophical and epistemological framework and hence they do not
seem to care about the existence or non-existence of an absolute frame
of reference and this should be endorsed by their technical and local
nature; see $\S$ \ref{secNonNewtonian}).\footnote{It is useful to note that the existence of a cosmological (or global)
absolute frame of reference does not necessarily mean the existence
(or non-existence) of a state of absolute rest and uniform rectilinear
motion. In fact, this should depend on the nature of this absolute
frame and its actual realization. For example, if the absolute frame
is realized through an infinite Newtonian-type frame (i.e. infinite
absolute space-time) then a state (of absolute rest and uniform rectilinear
motion) may not be possible to define sensibly. On the other hand
if the absolute frame is realized through a Machian-type frame then
such an absolute state may be defined sensibly. In fact, some of the
confusion and absurdity in special relativity may originate from the
lack of proper distinction between the existence of absolute frame
of reference and the existence of a state of absolute rest and uniform
rectilinear motion.}

In fact, the mere existence of the paradigms of inertial and non-inertial
frames in the Newtonian formulation is based on (and cannot be justified
without) an implicit assumption of the existence of a cosmological
(or global) absolute frame of reference. This is because any local
frame of reference should be ultimately referred to a global reference
frame at cosmological scale (i.e. by envisaging a hierarchical sequence
of reference frames which are ultimately referred to a single frame
at cosmological level). Accordingly, the existence of an absolute
frame of reference is not only a prior philosophical assumption (as
it is the case historically) but it is also a logical and epistemological
requirement for the Newtonian formulation of classical mechanics.
In other words, the presumption of the existence of an absolute frame
of reference is intrinsic to the formalism of Newtonian formulation
(see point \ref{enuIntrinsicExtrinsic} of $\S$ \ref{subsecConceptualFramework})
and not only a philosophical attachment that have been added to this
formulation for historical reasons.

In this regard we should draw the attention to the following remarks:
\begin{enumerate}
\item We use ``absolute frame'' as a term more general than the Newtonian
absolute space (as well as time) and the ``Machian frame'' which
is based on the overall distribution of matter in the Universe (according
to Mach's proposal about the origin of inertia). In fact, ``absolute
frame'' is more general than these two proposed absolute frames and
than any other potential or proposed absolute frames such as the Cosmic
Microwave Background Radiation (CMBR) or the luminiferous ether, i.e.
any one of these four potential or proposed frames of reference, as
well as any frame other than these four, can be accepted as realization
to the concept of ``absolute frame'' as long as it can produce sensible
and consistent epistemology (which should also be compliant with a
correct formalism that is based on this realization of ``absolute
frame''). Furthermore, we may accept more than one of these absolute
frames simultaneously as long as this condition is satisfied, i.e.
they can produce sensible and consistent epistemology which is in
compliance with a formalism based on these multiple realizations.\footnote{Accepting more than one absolute frame simultaneously may be collectively
and in combination (where each frame can provide a reference for certain
part of the formalism or/and epistemology of physics) or individually
and independently (where each frame can provide a reference for certain
formalism or/and epistemology for the entire physics). In fact, the
role of simultaneous multiple absolute frames can be envisaged in
many different ways, situations and scenarios (and this should be
inline with the principle of non-uniqueness of science and an instance
of it; see $\S$ 5.3 of \cite{SochiEpistIntro2024}). These issues
should be investigated in the future papers of this series.}
\item We should note that ``absolute frame'' includes (absolute) time
as well as (absolute) space and hence it is more general than the
aforementioned Newtonian absolute space (which is specifically about
space only) and ``Machian frame'' (which is apparently about space
only).
\end{enumerate}
\vspace{0.2cm}So in brief, ``absolute frame'' is more general from
these two perspectives, i.e. the perspective of various realizations
(i.e. the aforementioned four or more realizations) and the perspective
of inclusion of absolute time.

\subsubsection{\label{subsecCentralityForce}Centrality of ``Force'' Concept}

Here we discuss the fact that the central concept in the Newtonian
formulation of classical mechanics is the concept of force. This should
be obvious because this concept is common to all the three laws of
the Newtonian formulation and it is the subject of proposition in
Newton's second and third laws (as well as being implicitly so in
Newton's first law). It should be obvious that the concept of force
as a ``pull or push'' (or something like these) is intuitive although
its technical meaning may not be so (as we will highlight some of
its potential technical and epistemological ambiguities).

In fact, it may be claimed that there is an ``intrinsic ambiguity''
in the concept of force because there is no technical definition of
this concept (within the Newtonian formulation) independent of Newton's
laws (noting that these laws are based on this central concept and
hence using them as a basis for a technical definition may be seen
as a kind of circularity). However, this concern may be dismissed
by considering ``force'' as a basic term whose concept is defined
generically and elementarily by its intuitive meaning (and hence we
can be content with this type of ``definition'' since it is sufficient
for most or all theoretical and practical purposes formally and epistemologically).

Anyway, we should point out to the following potential ambiguities
(or similar difficulties) in or around the concept of force within
technical contexts especially from an epistemological viewpoint:
\begin{enumerate}
\item Ambiguity arising from the distinction (within the Newtonian formulation)
between fictitious and non-fictitious forces (see $\S$ \ref{subsecFictitiousForce}).
\item Ambiguity arising from the distinction (within the Newtonian formulation)
between action and reaction forces (see $\S$ \ref{subsecMissingAsymmetry}).\footnote{What we actually mean is the required and necessary distinction between
independent and dependent forces (which the Newtonian formulation
fails to do apart from a trivial labeling of ``action'' and ``reaction'').
This ambiguity should be clarified by reading $\S$ \ref{subsecMissingAsymmetry}.}
\item Ambiguity of the concept of ``inertia'' (which is a pivotal concept
in the Newtonian formulation) from a technical perspective noting
that the definition of this concept is usually based on (or associated
with) the concepts of force (which contains the aforementioned ambiguities)
and inertial forces.
\end{enumerate}
\vspace{0.15cm}These issues (and similar issues) should be investigated
directly or indirectly (or touched on) in the future.

\subsubsection{Additivity Principles}

A physical quantity (such as mass) is ``additive'' if this quantity
for a composite system is the sum of this quantity for its individual
components. For example, the mass of a two-particle system is the
sum of the masses of these two particles (unlike the temperature of
this system, for instance, which is not additive in this sense since
it cannot be obtained by summing the temperatures of the two particles).

The Newtonian formulation (and classical mechanics in general) contains
a number of additivity principles regarding certain additive physical
quantities (noting that most of these additivity principles are implicit).
These additivity principles (or rules) have theoretical justifications
(some of which are discussed in the literature of classical mechanics)
as well as experimental and observational verifications. In fact,
although these additivity principles rarely ``appear in public'',
they underlie most physical arguments and theoretical models in the
Newtonian formulation (and some even in the non-Newtonian formulations)
of classical mechanics and hence they play (implicit or unseen) fundamental
roles in the Newtonian formulation of classical mechanics (both formally
and epistemologically).

It is worth noting the following points about the additivity principles
in the Newtonian formulation:
\begin{enumerate}
\item ``Additivity'' here should include the additivity of scalar quantities
(such as the additivity of mass) and the additivity of vector quantities
(such as the additivity of force which represents superposition of
force vectors).
\item Some additivity principles may be linked to some conservation laws
(e.g. mass additivity and mass conservation); see $\S$ \ref{subsecConservationLaws}.
\item ``Composite system'' in the above definition of additivity can be
extended and generalized to include frames of reference in the sense
that a given physical quantity for a given object in a given frame
of reference (say frame A) is the sum of the value of that quantity
in another frame (say frame B) plus the value of that quantity that
characterizes the relationship between these frames. For example,
according to the Newtonian mechanics the velocity of a particle (in
a given direction) in frame A is the velocity of the particle in frame
B plus the velocity of frame B relative to frame A. Hence, the velocity
in the Newtonian mechanics is additive in this sense (which is not
the case in the relativistic mechanics for instance).
\end{enumerate}

\subsubsection{\label{subsecConservationLaws}Conservation Laws}

The Newtonian formulation of classical mechanics embeds a number of
conservation laws (or principles) which are:
\begin{enumerate}
\item The conservation of mass which is a postulated fundamental principle.
\item The conservation of mechanical energy (within certain conditions)
which can be derived from Newton's laws of motion (usually with the
requirement of some extra considerations and constraints and depending
on the given assumptions).
\item The conservation of (linear and angular) momentum (within certain
conditions) which can be derived from Newton's laws of motion (usually
with the requirement of some extra considerations and constraints
and depending on the given assumptions).
\end{enumerate}
\vspace{0.1cm}It is useful to note that these conservation laws (especially
those of energy and momentum) are not restricted to the Newtonian
formulation of classical mechanics (see $\S$ \ref{secNonNewtonian}).
Moreover, as indicated already the conservation laws of energy and
momentum are derivable (within certain conditions and assumptions)
from Newton's laws and hence they are not central to the formalism
of the Newtonian formulation although they (with the conservation
of mass) are epistemologically significant and can be central to the
epistemology and interpretation of classical mechanics in its Newtonian
formulation. In fact, (at least) some of these conservation laws are
physically more fundamental than some of Newton's laws (and hence
when we say ``they are not central'' it does not mean that their
physical content is marginal or less significant but it means that
they are derivable from Newton's laws which represent the main and
primary body of the Newtonian formulation).

We should also note that the views and methods used in the literature
to present and establish the above conservation laws vary in general,
and hence the above statements about these laws (e.g. whether they
are postulated or derived and how and why) represent our views (or
rather certain views) which may contradict other views in the literature.
However, these details are marginally significant to our current investigation
and purposes and hence we ignore these details (referring the interested
readers to the literature of classical mechanics).

\subsection{Limitations of Newtonian Formulation of Classical Mechanics}

The main epistemological limitations and shortcomings (which are largely
related to interpretability and interpretation) of the Newtonian formulation
of classical mechanics are investigated briefly in the following subsections
(noting that some of these limitations and shortcomings are related
to the formalism of the Newtonian formulation as well although this
is of no interest to us in this investigation whose focus is epistemology).
More specific and detailed investigations to some epistemological
limitations and shortcomings of this formulation should be pursued
in forthcoming papers of this series.

\subsubsection{\label{subsecLackOfDefinition}Lack of Technical Definition to Basic
Concepts}

The Newtonian formulation of classical mechanics may be criticized
by the absence of rigorous technical definitions to some of its basic
and fundamental concepts. For example, ``mass'' is usually defined
as the ``quantity of matter'' but this is not a rigorous technical
definition. Other proposed definitions of mass (which are supposedly
technical definitions) are also problematic such as being circular
(like its definition as the product of volume times density, i.e.
``\textit{mass} density'') or being dependent on Newton's laws (as
it is the case with the definition based on the reciprocal proportionality
to the magnitude of accelerations which is essentially no more than
a ``rumination'' to Newton's second law or a sort of tautology noting
that Newton's second law itself requires the definition of its basic
concepts and ingredients in advance to be completely and technically
determined and hence it cannot be used in this way as a basis for
a technical definition to some of its concepts or ingredients).

This also applies to ``force'' which is usually defined as a ``push
or pull'' (which is not technical) or by Newton's second law (which,
as before, requires the definition of its basic concepts and ingredients
in advance to be completely and technically determined and hence it
cannot be used in this way as a basis for a technical definition to
some of its concepts or ingredients).

However, the lack of technical definitions may not be seen as a serious
problem from an epistemological perspective (which is our focus in
this investigation) noting the intuitivity of most of the elements
and ingredients of the conceptual framework of the Newtonian formulation
(as indicated in $\S$ \ref{subsecConceptualFramework}). Nevertheless,
regardless of accepting or rejecting this defence the lack of rigorous
technical definitions to some basic concepts and ingredients (independently
of the formulation itself) should be considered as a shortcoming in
the sense that it is better (at the least) to have rigorous technical
definitions to all the basic concepts and ingredients of this formulation
so that the formulation be completely technical and rigorous (and
this is particularly important to the epistemology and interpretation
of this formulation).

We should also remember in this context the ambiguities surrounding
the concept of force (as discussed in $\S$ \ref{subsecCentralityForce})
noting that some of these ambiguities are not related (at least directly
and primarily) to the definition of the concept of force. These ambiguities
(and potentially other ambiguities) should cast a shadow on the clarity
and integrity of any epistemology and interpretation of the Newtonian
formulation (and some even on classical mechanics in general).

\subsubsection{\label{subsecRestrictedInertial}Restricted Validity to Inertial
Frames}

One of the main limitations of the Newtonian formulation of classical
mechanics is that the validity of this formulation is restricted to
inertial frames. This restriction on the validity of the formalism
of this formulation should impose a parallel restriction on the validity
and extension of its epistemology and interpretation. Yes, the validity
of this formulation may be extended (in some sense) to non-inertial
frames by the introduction of the concept of ``fictitious force''
which supposedly compensates for the failure of this formulation in
non-inertial frames. However, the concept of ``fictitious force''
is not free of problems (at least from an epistemological perspective;
see $\S$ \ref{subsecFictitiousForce}), and hence this remedy does
not seem to address the original problem entirely and satisfactorily.

We should also note that the concept of ``inertial frame'' may be
seen as problematic theoretically\footnote{For instance, it may be claimed that ``inertial frame'' cannot be
defined and identified by the validity of Newton's laws (as it is
commonly done in the literature) due to circularity (or potential
circularity).} (and even practically), and this should add another source of epistemological
(and even formal) difficulties to this formulation from this perspective.
However, we think the acceptance of the paradigm of ``absolute frame''
(which is one of the main features of the Newtonian formulation; see
$\S$ \ref{subsecAbsoluteFrame}) regardless of the physical realization
and origin of this frame (whether by Newton's absolute space and time
or by Machian-type frame or by the CMBR or by the ether or by something
else) should address this issue satisfactorily (at least theoretically
and epistemologically).

\subsubsection{\label{subsecRestrictedScale}Restricted Validity to Classical Macroscopic
Scale}

We should remind the reader first that ``scale'' is more general
than ``size'' (see $\S$ 7.2 of \cite{SochiEpistIntro2024}). The
following are two obvious and famous examples of the restrictions
on the validity of the Newtonian formulation of classical mechanics
due to limitations imposed by the ``scale'' factor:
\begin{enumerate}
\item From the perspective of the ``size'' scale, the Newtonian formulation
is invalid (at least in its basic form) at quantum and sub-quantum
scales. In our view, it should also be invalid at cosmological scale.
Its validity at astronomical scale is at least tentative (noting that
there seems to be indications of its invalidity at this scale). So
in brief, the validity of the Newtonian formulation is certainly restricted
(epistemologically as well as formally) to certain size scale(s).
\item From the perspective of the ``speed'' scale, the Newtonian formulation
is similarly invalid (at least in its basic form) at high speeds (i.e.
comparable to the speed of light) and this is what necessitates the
introduction of Lorentz transformations and mechanics (as well as
other proposed physical theories that deal with this problem regardless
of accepting or rejecting Lorentz transformations and these proposed
theories). So, the validity of the Newtonian formulation is restricted
(epistemologically as well as formally) to certain speed scale(s).
\end{enumerate}
\vspace{0.15cm}In fact, other potential limitations on the Newtonian
formulation due to scale restrictions (from perspectives other than
size and speed) are also possible. Anyway, regardless of these (and
other related) details the validity of the Newtonian formulation of
classical mechanics is certainly restricted (formally and epistemologically)
to certain \textit{scales} and hence it cannot provide a full epistemological
theory or interpretation even if we ignore or address other limitations
and shortcomings of this formulation.

It is worth noting that it is not enough to address this problem (as
well as the previous problem; see $\S$ \ref{subsecRestrictedInertial})
by considering or labeling the Newtonian formulation (or even the
entire classical mechanics) as an approximation to a more fundamental
theory. In other words, this is not a solution to this limitation
(in fact this ``solution'' is essentially an admission of this limitation
as well as other similar limitations).

\subsubsection{\label{subsecFictitiousForce}Ambiguity of Origin of Fictitious Force}

The concept of ``fictitious force'' in the Newtonian formulation
of classical mechanics was invented to account for the failure of
Newton's laws in non-inertial frames of reference. To put it in more
friendly terms, this concept was introduced to the Newtonian formulation
to ``extend the validity'' of Newton's laws from inertial frames
of reference to non-inertial frames of reference. Accordingly, ``fictitious
force'' may be defined as a hypothetical force that should be assumed
to exist in a non-inertial frame of reference to make Newton's laws
of motion applicable in that frame.

Anyway, we should note that fictitious forces are real forces because
they have observable physical effects (such as the effects of Coriolis
force on the global winds and ocean currents) and they can be directly
felt by our bare senses (such as the feeling of pressure on our muscles
and organs when traveling in an accelerated vehicle or boarding a
taking-off plane). So, ``fictitious'' is actually a technical term
to indicate that these forces are restricted to non-inertial frames
(i.e. they do not exist in inertial frames) and hence ``fictitious''
should not be understood to mean something like ``imaginary'' or
``illusory''.\footnote{In fact, they are ``imaginary'' and ``illusory'' from a Newtonian
viewpoint and in a certain sense because (as we will see) they have
no known physical origin or observable genesis within the conceptual
framework of the Newtonian formulation. We may also rationalize the
``fictitious'' nature of these forces by saying that an observer
in a non-inertial frame cannot identify a physical origin or tangible
source to them and hence in his view they are imaginary and illusory
(or ``fictitious'').}

What distinguishes fictitious forces from ``real'' forces is that
they do not have a directly observable origin.\footnote{For instance, we read in French (see page 509 of \cite{FrenchBook1971}):
The observer explains the extension of the spring by saying that it
is counteracting the outward centrifugal force on $m$ which is present
in the rotating frame. Furthermore, if the spring breaks, then the
net force on the mass is just the centrifugal force and the object
will at that instant have an outward acceleration of $\omega^{2}r$
in response to this so-called ``fictitious'' force. Once again the
inertial force is ``there'' by every criterion we can apply (except
our inability to find another physical system as its source). (End
of quote)} For example, the centripetal force is ``real'' because it is exerted
by directly observable things like a string (connected to a rotating
ball) or magnetic attractive field or gravitating massive body, but
the centrifugal force is ``fictitious'' because there is no directly
observable thing to which it can be attributed although in reality
there should be some origin to it such as being an intrinsic property
of space\footnote{We may distinguish between \textit{geometric space} which is an abstract
mathematical entity and \textit{physical space} which (supposedly)
is a real physical entity with certain properties and attributes such
as having electric permittivity and magnetic permeability (which determine
the speed of light in \textit{vacuum} for instance) or exerting certain
gravitational and inertial influences and effects (and so on). In
fact, the paradigm of physical space in this sense is acceptable formally
and epistemologically as long as it serves a legitimate role and a
justified purpose in a consistent physical theory.} or being the result of overall gravity of the matter in the Universe
(as may be explained by Mach's proposal for instance).

From our viewpoint, the concept of ``fictitious force'' should be
seen as an epistemological defect or gap in the Newtonian formulation
of classical mechanics because since it is a real force\footnote{As indicated earlier, fictitious forces are real forces in commonsense
and according to our direct sensual experiences although they are
not real in a technical meaning according to the Newtonian formulation
of classical mechanics.} it should have a physical origin and authentic genesis (according
to the principle of causality; see $\S$ 5.2 of \cite{SochiEpistIntro2024})
and hence the absence of this origin in the theory creates a gap (at
least epistemologically) in this theory. It is worth noting that the
concept of ``fictitious force'' does not exist in the other two
main formulations of classical mechanics (namely the Lagrangian and
Hamiltonian mechanics) because they, unlike the Newtonian formulation,
are not based or centered on the concept of ``force'' (see $\S$
\ref{secNonNewtonian}).

\subsubsection{\label{subsecViolationThirdLaw}Violations of Newton's Third Law}

There are discussions and investigations in the literature of mechanics
and electromagnetism about violations of Newton's third law in some
physical phenomena (or theories or branches of physics). More specifically,
we refer to the claimed violations of Newton's third law (i.e. from
a formal perspective) in electrodynamics and relativistic mechanics
(as well as other potential violations) which are discussed in the
literature.

Accordingly, any potential or proposed epistemology of the Newtonian
formulation to classical mechanics should address this \textit{defect
in the formalism} of this formulation which should obviously have
an impact on its epistemology and interpretability or interpretation.
In fact, this is a big issue and hence we postpone the discussion
and investigation about it to the future papers of this series.

\subsubsection{\label{subsecMissingAsymmetry}Missing Asymmetry in Newton's Third
Law}

It is obvious that from a purely technical and formal perspective
there is an implicit symmetry in Newton's third law (i.e. the two
forces in this law are represented symmetrically without any technical
or formal distinction between them noting that the sign is arbitrary).
This symmetry is misleading or not describing the actual physical
situation completely and clearly (noting that the physical situation
is actually asymmetric at least in some cases). In other words, the
implicitly-presumed ``formal'' symmetry of forces in Newton's third
law is not sufficient or able to explain the physical situation epistemologically
(at the least) in these circumstances because (assuming an interaction
between agent A and agent B) we can easily distinguish between the
situation when A pushes B and the situation when B pushes A although
from a purely formal perspective of Newton's third law the two situations
are identical. In fact, we can see this distinction from a causality
perspective where in one situation A is the cause of interaction while
in the other situation B is the cause of interaction.

The mere labeling of one force as action and the other force as reaction
(as it is usually done in the statement of Newton's third law) is
not enough to make a clear and realistic physical and epistemological
distinction between the two situations (noting as well that this seeming
distinction by labeling is external and is not included or embedded
technically within the formalism or epistemology of the physical situation).
In other words, we must have a hidden (real and physical) element
or ingredient in these physical situations that distinguishes the
two cases, and this element is not incorporated within the theory
in its formal and technical aspect (and hence it is essentially missing
in its epistemological aspect noting as well that there is no such
distinction within the conceptual framework extrinsically). The failure
to include this element makes the theory ambiguous epistemologically
(as well as possibly being formally incomplete).

To represent the physical situation (in the above example and its
alike) completely and clearly, the law (or an extension or attachment
to the law) should distinguish realistically, technically and intrinsically
between two types of force: independent force (which is usually labeled
as ``action'') and dependent force (which is usually labeled as
``reaction''). It should be obvious that there is a real physical
difference between these types of forces (noting that the emergence
of the dependent force is caused by the existence of the independent
force but not the other way around or symmetrically) and this difference
is neither incorporated within Newton's third law formally and technically
nor identified or clarified epistemologically (i.e. what is the physical
element or ingredient that qualifies a force to be an independent
action force while another force as a dependent reaction force?).
So in our view, if this missing asymmetry is not a gap in the Newtonian
formulation from the perspective of formalism, it should be at least
a gap in this formulation from the perspective of epistemology.

We should finally note that in some physical circumstances and instances
the physical situation is seemingly symmetric, e.g. agent A and agent
B push each other equally and symmetrically. However, this should
not affect the fact that the action (or independent) force and the
reaction (or dependent) force in these circumstances are still asymmetric
because what is symmetric in these circumstances is the two action
forces of the two agents (as well as the two reaction forces) and
hence the action force of each agent and its corresponding reaction
force are still asymmetric.\footnote{In fact, a closer look should reveal that we actually have (at least)
three distinct cases:\\
1. A person pushes a wall: the situation in this case is obviously
asymmetric because the independent action belongs to the person while
the dependent reaction belongs to the wall.\\
2. Two gravitating objects (e.g. Sun and Earth) or two attracting/repelling
electric charges: in this case it seems we have a symmetric situation
since the force exerted by each object or charge can be regarded as
action or reaction (although we may assume that we have two action
forces and two reaction forces, but this seems to complicate the situation
unnecessarily).\\
3. Two persons push each other equally and symmetrically: in this
case we should assume (to be logical) that we have two action forces
and two reaction forces (and this case is what we indicated in the
main text above).\\
So, each one of these situations requires investigation and distinction
at the epistemological level to identify what physical factor distinguishes
the situation in each one of these cases. Anyway, the mere existence
of these three distinct cases should support our proposal that Newton's
third law does not depict the physical situation completely (at least
from epistemological perspective and in some cases) and this can be
seen as a gap in the epistemology of the Newtonian formulation.}

\subsection{Interpretability and Interpretation of Newtonian Formulation}

Based on our investigation in the previous three subsections we can
conclude that although the Newtonian formulation of classical mechanics
possesses a rather elaborate conceptual, philosophical and epistemological
framework which qualifies this formulation to be interpretable \textit{in
principle}, this framework lacks sufficient epistemological consistency
and integrity to be a proper basis for consistent and complete interpretation
as it stands.

This means, in short, that the Newtonian formulation is interpretable
in principle but does not actually have an interpretation. Yes, if
certain shortcomings and defects in this formulation are addressed
and corrected it may become eligible for having a proper interpretation
(i.e. within its limited domain of validity or even within larger
domain of validity depending on the proposed modifications, remedies
and rectifications).

So in brief, although the recipe of the Newtonian formulation represents
a good and successful formalism at the practical level (within its
domain of validity and the required conditions and assumptions) it
does not have a sensible and complete epistemology or a valid and
consistent interpretation due to these problematic issues.

\section{\label{secNonNewtonian}Non-Newtonian Formulations of Classical Mechanics}

The non-Newtonian formulations of classical mechanics generally come
under the title of ``analytical mechanics''. The most common of
these formulations are the Lagrangian and Hamiltonian mechanics. It
is worth noting the following points about the non-Newtonian formulations
of classical mechanics:\footnote{In fact, these points are supposed to provide a brief investigation
to the non-Newtonian formulations of classical mechanics in general
where we assess their epistemological merit and their potential interpretation(s)
or even their interpretability. More detailed and specific investigations
about these formulations and related issues may be pursued in the
future.}
\begin{enumerate}
\item These formulations have very mathematical and technical nature and
hence they should be classified more appropriately as branches of
mathematical physics rather than physics. In other words, these formulations
are essentially mathematical methods for classical mechanics rather
than independent paradigms (with independent and specific conceptual,
philosophical and epistemological frameworks) of classical mechanics,
and hence from this perspective they do not contrast the Newtonian
formulation of classical mechanics (which has elaborate, independent
and specific conceptual, philosophical and epistemological framework).
Accordingly, the central focus of these formulations is formalism
not epistemology and interpretation since they are not associated
with proper conceptual, philosophical and epistemological frameworks.
So, in this regard they may be likened to quantum mechanics which
(in our view) is a formalism without a sensible epistemology or a
valid interpretation, i.e. it is not interpretable (see $\S$ 8.6
of \cite{SochiBook8}).
\item The fact that the Newtonian formulation is vector-based while the
non-Newtonian formulations are scalar-based may grant these formulations
more ``coordinate-flexibility'', and this may be seen as a sign
or indication of the specific characteristic of the Newtonian space
(which is supposedly Euclidean) in contrast to the space that these
formulations can represent and apply to (i.e. they, unlike the Newtonian
formulation, have ``space-flexibility'' corresponding to their ``coordinate-flexibility''
in the sense that they can in principle work in Euclidean and non-Euclidean
spaces if such flexibility is required). This may be seen as an advantage
to these non-Newtonian formulations that can have not only formal
benefits but also epistemological benefits.
\item These formulations (or at least some of them) have limitations in
validity and application, e.g. they may not work in certain physical
systems like non-conservative systems.\footnote{We are referring in this point to limitations that are usually not
found in other formulations of classical mechanics (particularly the
Newtonian formulation), and hence they represent extra limitations.} Hence, even if we assume that they have epistemology and interpretation
they are limited in these aspects following their limitation in validity
and application.
\item These formulations (like the Newtonian formulation; see $\S$ \ref{subsecRestrictedScale})
are scale-limited since their validity is generally restricted (at
least tentatively) to the classical macroscopic scale (although some
of these formulations have adaptations and variations used in other
scales, e.g. in dealing with quantum or cosmological systems).
\item (At least) some of these formulations are based on (or formulated
in) non-intuitive abstract concepts and principles (like the concept
of ``action'' and ``the principle of least action'') and this
should make any \textit{potential} epistemology and interpretation
to these formulations more abstract and difficult (or rather ``more
impossible'' considering other factors).\footnote{In fact, even the concept of ``energy'' (which is the basis of some
non-Newtonian formulations and which is less abstract than the concept
of ``action'') is more abstract (and hence less intuitive) than
the concept of force (which is the central concept in the Newtonian
formulation) and this should make any interpretation more difficult
(noting that interpretation is essentially based on intuition).}
\item It is shown in the literature that these non-Newtonian formulations
are equivalent to the Newtonian formulation in (at least some) formal
aspects (e.g. Newton's laws can be derived from the Lagrangian and
Hamiltonian formulations) and this may be seen as a basis for these
non-Newtonian formulations to have (in principle) proper epistemology
and interpretation due to this equivalence to the Newtonian formulation.\footnote{In fact, there are many aspects (and potential interpretations) to
this supposed equivalence which require extensive investigation and
inspection. We may discuss some of these aspects in the future.} However, this formal equivalence should not make these non-Newtonian
formulations epistemologically equivalent to the Newtonian formulation
(because formal equivalence does not imply epistemological equivalence).
Nevertheless, even if an epistemology of these non-Newtonian formulations
is obtained (presumably) through their equivalence to the Newtonian
formulation, the obtained epistemology is actually the epistemology
of the Newtonian formulation and hence it should not be credited to
these non-Newtonian formulations (regardless of the obtained epistemology
being similar or dissimilar to the Newtonian epistemology).
\end{enumerate}
\vspace{0.2cm}So in brief, while the Newtonian formulation of classical
mechanics has a rather clear and elaborate conceptual, philosophical
and epistemological framework, the non-Newtonian formulations (like
the Lagrangian and Hamiltonian mechanics) do not possess such a framework.
Hence, the Newtonian formulation is interpretable in principle (even
though it may not have a proper, complete and consistent interpretation
due to its epistemological shortcomings, defects and gaps as outlined
earlier) while the non-Newtonian formulations are not interpretable
due to the aforementioned lack of proper conceptual frameworks.

\clearpage{}

\section{Conclusions}

We outline in the following points the main facts and results that
we discussed and obtained in this paper:

\noindent $\bullet$ While the Newtonian formulation of classical
mechanics possesses a rather elaborate and intuitive conceptual, philosophical
and epistemological framework which qualifies it (in principle) to
be interpretable and have a proper interpretation, the non-Newtonian
formulations of classical mechanics do not possess such a framework.

\noindent $\bullet$ Although the Newtonian formulation of classical
mechanics is interpretable in principle (thanks to its aforementioned
framework) it does not have a proper epistemology or consistent and
complete interpretation due to epistemological defects and gaps. On
the other hand, the non-Newtonian formulations of classical mechanics
are not interpretable due to the lack of conceptual, philosophical
and epistemological frameworks that are required for any interpretation,
and hence these formulations can be regarded as formalism without
an epistemology that qualifies them to have an interpretation in principle
(i.e. they are like quantum mechanics in this regard).

\noindent $\bullet$ To conclude, the Newtonian formulation has no
sensible interpretation, while the non-Newtonian formulations are
not interpretable. Hence, none of the formulations of classical mechanics
have a proper epistemology or sensible interpretation.

\clearpage{}

\phantomsection 
\addcontentsline{toc}{section}{References}\bibliographystyle{unsrt}
\bibliography{Bibl}

\end{document}